\documentclass{article}
\pdfoutput=1

\usepackage{monash}

\bibliography{DM}

\title{Halo-independence with quantified maximum entropy at DAMA/LIBRA}
\author{Andrew Fowlie}
\affiliation{ARC Centre of Excellence for Particle Physics at the Tera-scale, Monash University, Melbourne, Victoria 3800, Australia}

\begin{document}

\maketitle

\begin{abstract}
Using the DAMA/LIBRA anomaly as an example, we formalise the notion of halo-independence in the context of Bayesian statistics and quantified maximum entropy. We consider an infinite set of possible profiles, weighted by an entropic prior and constrained by a likelihood describing noisy measurements of modulated moments by DAMA/LIBRA. Assuming an isotropic dark matter (DM) profile in the galactic rest frame, we find the most plausible DM profiles and predictions for unmodulated signal rates at DAMA/LIBRA. The entropic prior contains an a priori unknown regularisation factor, $\beta$, that describes the strength of our conviction that the profile is approximately Maxwellian. By varying $\beta$, we smoothly interpolate between a halo-independent and a halo-dependent analysis, thus exploring the impact of prior information about the DM profile.
\end{abstract}

\section{Introduction}

The DAMA/LIBRA search for dark matter (DM) saw an intriguing modulated signal above known backgrounds\cite{Bernabei:2008yi,Bernabei:2010mq,Bernabei:2013xsa}. Although the statistical significance of the effect, about $9\sigma$, is beyond doubt, there is controversy about whether the signal is consistent with competing experiments such as XENON\cite{Aprile:2017yea,Aprile:2017iyp} and LUX\cite{Akerib:2016lao}, and whether it is an effect of DM or an unidentified seasonal systematic error. The velocity profile of DM is an important component of any DM explanation of the DAMA/LIBRA anomaly. We expect that it is approximately a truncated Maxwell-Boltzmann distribution,
\begin{equation}\label{eq:mb}
m(v) \propto \begin{cases} 
        v^2 e^{-\left(\frac{v}{v_0}\right)^2} & v < v_\text{esc}\\
        0 & v \ge v_\text{esc}
        \end{cases},
\end{equation}
where  $v_\text{esc}$ and $v_0$ are the escape and modal velocities, respectively, and $\int_{\mathbf v} m(v) dv = 1$. This is consistent with a self-gravitating isothermal sphere with a  density profile $\rho(r) \propto 1 /r^2$ and agrees reasonably with $N$-body DM only simulations\cite{Kuhlen:2009vh}. In the last 10 years, an effort was made to analyse direct detection experiments, including DAMA/LIBRA, in a so-called halo-independent manner by isolating the impact of the DM velocity profile and making no restrictive assumptions about it \see{HerreroGarcia:2012fu,DelNobile:2013cta,DelNobile:2013cva,Gondolo:2012rs,Fox:2014kua,Feldstein:2014gza}. 

The DAMA/LIBRA experiment reported counts with uncertainties for a modulated signal per unit detector mass per day binned with respect to energy (see \reftable{tab:sm}). The counts from a DM model can be predicted from moments of the velocity profile. As recognised in \refcite{Gondolo:2017jro}, the problem is essentially a noisy moment problem: we wish to find a profile $f(v)$ that is consistent with DAMA/LIBRA measurements of its moments,
\begin{equation}
\int_{\mathbf v} f(v) w_i(v) dv = \mu_i \pm \sigma_i,
\end{equation}
where $w_i(v)$ are the moment functions and $\mu_i \pm \sigma_i$ are their measurements and experimental errors. Since the interval $\mathbf v$ is bounded by zero and the escape velocity, the problem is related to the Hausdorff moment problem, and would be the Hausdorff moment problem if the moment functions were $w_i(v) = v^i$.

An exact solution to the moment problem may not exist and may not be unique. The Pad\'e approximation is a common technique in which one expresses the density as a finite sum of Dirac delta functions,
\begin{equation}
f(v) = \sum_i \kappa_i \delta(v - v_i),    
\end{equation}
and solves for the unknown parameters $\kappa_i$ and $v_i$. This represents the DM velocity profile as a sum of discrete streams. Although this is an a priori implausible description of the profile, it may be used to estimate further moments. The Pad\'e approximation in fact extremises further moments, which was utilised in \refcite{Gondolo:2017jro} in the context of a frequentist analysis of DAMA/LIBRA and in \refcite{Gelmini:2017aqe,Ibarra:2017mzt}. That approach is halo-independent since it extremises further moments upon an infinite set of profiles, permitting the calculation of confidence intervals for e.g., the unmodulated signal rates at DAMA/LIBRA.

We, on the other hand, advocate investigating the impact of background knowledge about the profile with a non-parametric Bayesian analysis. In fact, by altering a regularisation factor in an entropic prior, we may smoothly interpolate between a halo-independent and halo-dependent analysis. To elaborate further in \refsec{sec:maxent} we first introduce pertinent aspects of Bayesian statistics and the principle of quantified maximum entropy. Readers familiar with quantified maximum entropy should skip to \refsec{sec:results}, in which we present our maximum entropy DM profiles. We conclude in \refsec{sec:conc}.

\section{Bayesian statistics and quantified maximum entropy}\label{sec:maxent}

We wish to apply Bayes' theorem in the form,
\begin{equation}\label{eq:bayes}
p(f\given \text{data, prior knowledge}) \propto
p(\text{data}\given f) \cdot p(f\given \text{prior knowledge}),
\end{equation}
to infer the most plausible DM velocity profile $f$ in light of our prior knowledge, favouring a truncated Maxwell-Boltzmann distribution, and noisy measurements of its moments by DAMA/LIBRA, under the assumption that the profile is isotropic in the galactic rest frame. The principle of maximum entropy was proposed by Jaynes in 1957\cite{PhysRev.106.620} by building upon Shannon's information theory\cite{BLTJ:BLTJ1338}. Just as Bayes' theorem is a unique law that satisfies Cox's axioms, the relative entropy generalised to continuous distributions,\footnote{When applied to the DM profile, \refeq{eq:entropy} implicitly assumes isotropy. Without that assumption, we would have to compute a three-dimensional integral for the entropy,
\begin{equation*}
S[f, m] = - \int f(\vec v) \ln \left(\frac{f(\vec v)}{m(\vec v)}\right) d^3v.
\end{equation*}
}
\begin{equation}\label{eq:entropy}
S[f, m] = - \int f(v) \ln \left(\frac{f(v)}{m(v)}\right) dv
\end{equation}
is a unique quantity that satisfies desiderata about a measure of information in a distribution $f(v)$ relative to a distribution $m(v)$. By Gibbs' inequality, $S[f, m] \le 0$, which is saturated when $f(v) = m(v)$.

To apply the principle of \emph{maximum entropy}, one finds a distribution $f(v)$ that maximises the entropy in \refeq{eq:entropy} subject to moment constraints on $f(v)$,
\begin{equation}
\int_{\mathbf v} f(v) w_i(v) dv = \mu_i.
\end{equation}
By using Lagrange multipliers, it can be shown that the solution is of the form \see{Mead},
\begin{equation}\label{eq:me}
f(v) \propto m(v) \exp\left(\sum_i \lambda_i w_i(v)\right).
\end{equation}
% The Lagrange multipliers $\lambda_i$ are typically solved by constructing a dual function, which changes the problem from constrained optimisation to unconstrained optimisation. The maximum entropy solution is unique.
The relative entropy is a strictly convex function such that the maximum entropy solution is unique.

There is controversy about the relationship between maximum entropy and Bayesian statistics \see{Jaynes1988, doi:10.1063/1.4819977}. We, in fact, favour constructing an entropic prior that penalises deviations from our prior knowledge $m(v)$. This approach --- known as \emph{quantified maximum entropy} --- was developed by Skilling\cite{skilling}. Our prior must be a monotonic function of the relative entropy between the profile and a truncated Maxwell-Boltzmann. As shown by Skilling's ``monkey'' argument\cite{skilling}, we find that
\begin{equation}
p(f\given \text{prior knowledge}) \propto e^{\beta S[f, m]}.
\end{equation}
The regularisation parameter $\beta$ describes the strength of our prior information: $\beta = 0$ corresponds to no prior information, whereas $\beta\to\infty$ is so strong that $f(v) = m(v)$ regardless of any other information or measurements. This entropic prior favours minimal information in the profile relative to background knowledge (see Chap.~12 of \refcite{linden_dose_toussaint_2014} for a review). 

% We might, in principle, wish to encode prior information about correlations in the profile. This could be achieved by expressing the profile as a convolution of an uncorrelated profile with an intrinsic correlation function, 
% \begin{equation}
% f(v) = \int C(v, v^\prime) h(v^\prime) dv^\prime
% \end{equation}
% where $h$ is uncorrelated and $C$ describes anticipated correlations. We could, alternatively, derive new expressions for the entropic prior that reflected correlations.

In the noisy moment problem, the posterior for the profile in \refeq{eq:bayes} contains a factor from our prior knowledge and a chi-squared factor from the noisy moment measurements;
\begin{equation}\label{eq:pdf}
p(f\given \text{data, prior knowledge}) \propto e^{\beta S[f, m] -\frac12\chi^2[f]},
\end{equation}
where
\begin{equation}
\chi^2[f] = \sum_i \left(\frac{\int_{\mathbf v} f(v) w_i(v) dv - \mu_i}{\sigma_i}\right)^2.
\end{equation}
The sum of the entropy and the chi-squared is a strictly convex function. If there is a stationary point, it is unique. As discussed in \refcite{Ciulli,Kopec1993}, there is a trade-off between penalties from the entropic prior for deviating from the prior information and penalties from the chi-squared for deviating from the measured moments. We can interpret $\beta$ by noting that an $n$ nat (natural unit of information) departure from $m(x)$  is punished by a factor equivalent to a chi-squared of $2\beta n$. The change in results with $\beta$ is known as a \emph{maximum entropy trajectory}. 

In the limit in which experimental uncertainties vanish, $\sigma \to 0$, the mode in this distribution is the maximum entropy distribution. In the general case, mode is of the form\footnote{We note that the maximum entropy profile resembles parametric forms of profile considered in Eq.~4 of \refcite{Kavanagh:2013wba}.}
\begin{equation}\label{eq:sol}
f(v) \propto m(v) \exp\left(\sum_i \kappa_i w_i(v)\right),
\end{equation}
where
\begin{equation}
\int_{\mathbf v} f(v) w_i(v) dv = \mu_i - \beta \kappa_i \sigma_i^2.
\end{equation}
As noted in \refcite{Kopec1993}, the coefficients $\kappa_i$ play the role of the Lagrange multipliers in \refeq{eq:me} for modified constraints. The maximum entropy solution in the noiseless moment problem is thus approximately the mode in the noisy moment problem when $\sigma \approx 0$. The frequentist analysis in \refcite{Gondolo:2017jro} roughly corresponds to $\beta \to 0$. Though for $\beta = 0$ the solution is not unique since the chi-squared is not strictly convex, we find the solution of the form in \refeq{eq:sol}, which is unique. We could, in principle, marginalise $\beta$ \see{gull1999quantified}.

\begin{figure}
    \centering
    \begin{subfigure}[t]{0.31\textwidth}
        \centering
        \includegraphics[width=\textwidth]{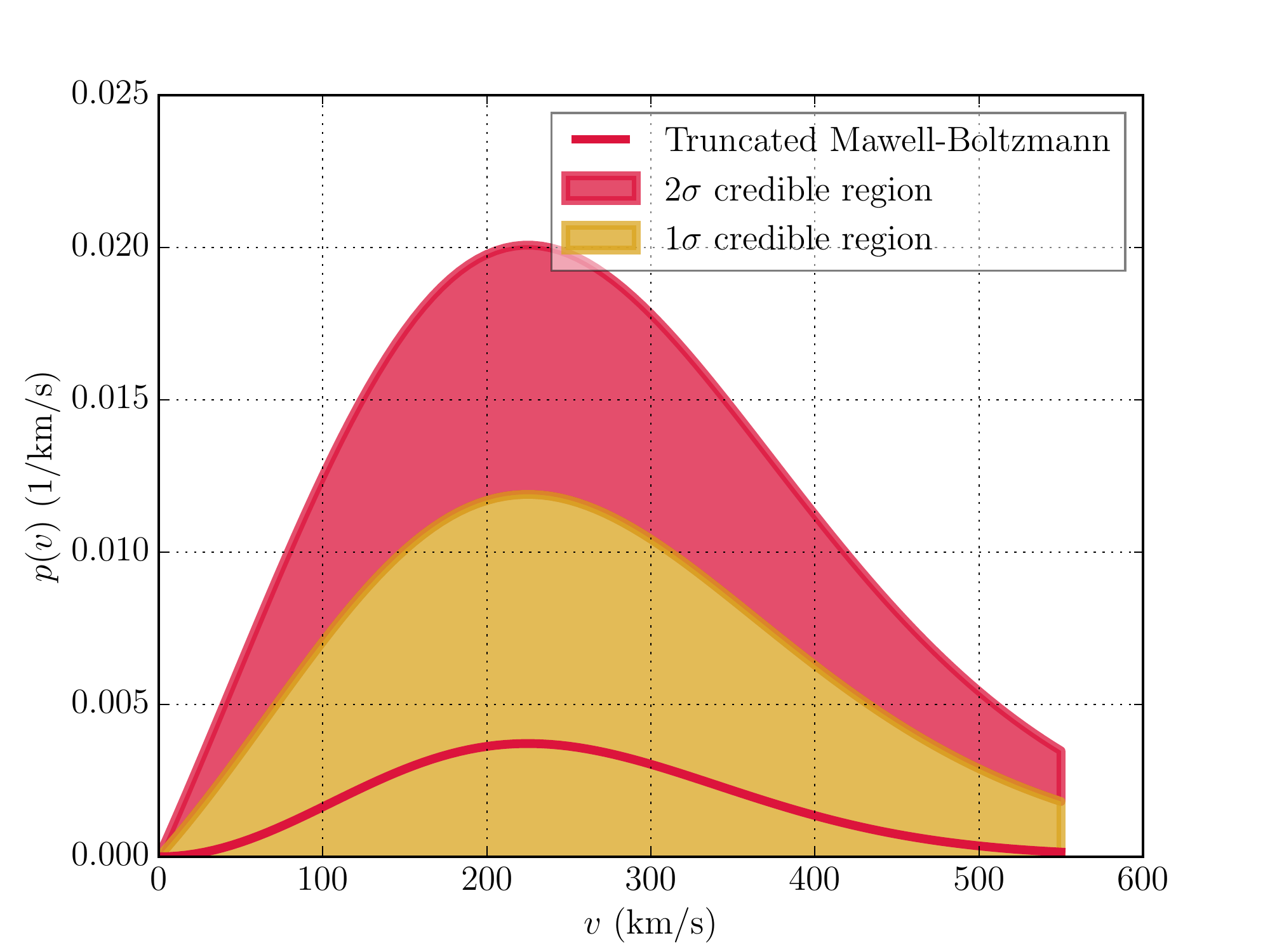}
        \caption{$\beta = 10$.}
        \label{fig:mistrust}
    \end{subfigure}
    \begin{subfigure}[t]{0.31\textwidth}
        \centering
        \includegraphics[width=\textwidth]{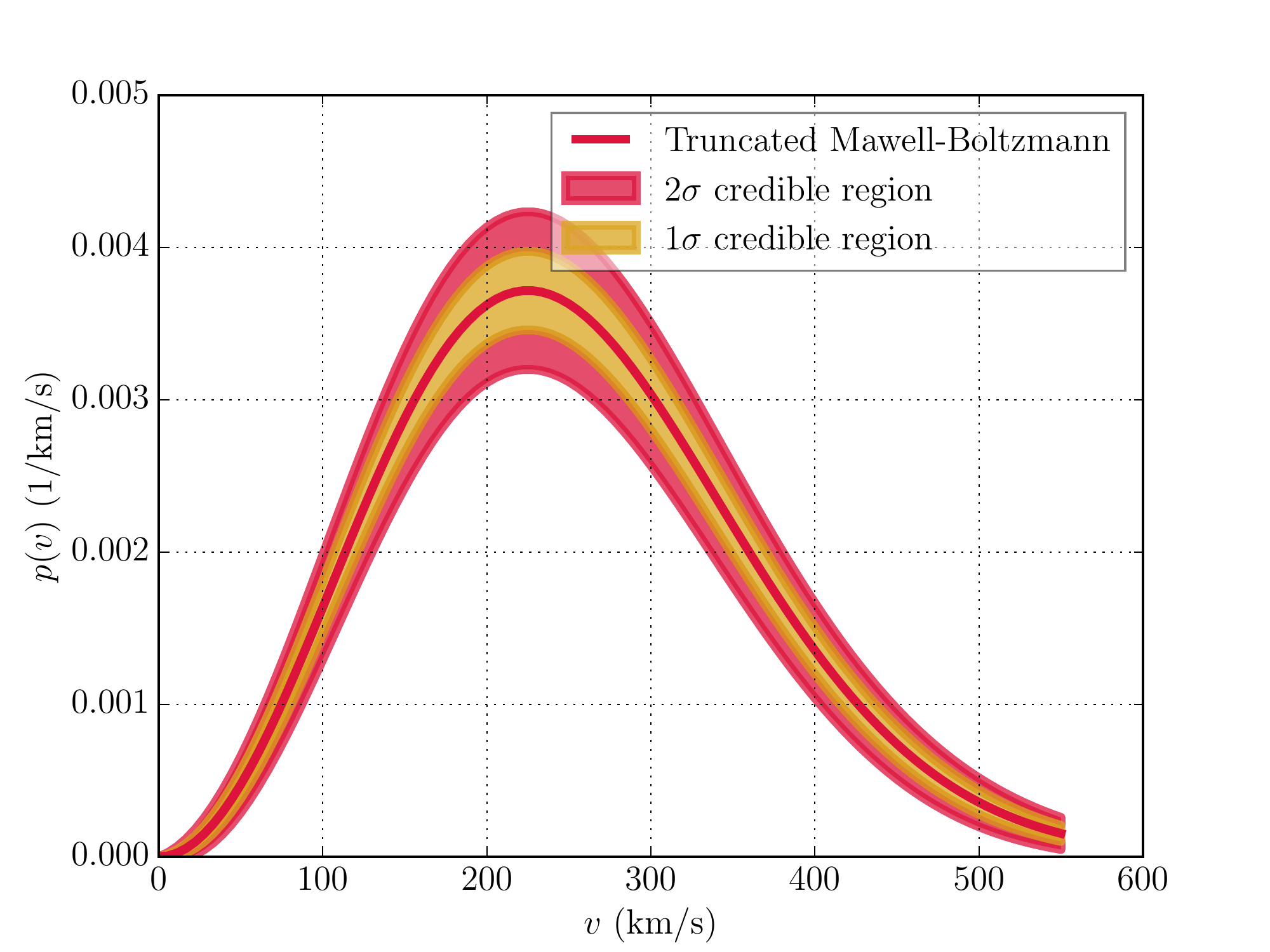}
        \caption{$\beta = 10^4$.}
        \label{fig:calibrated}
    \end{subfigure}  
    \begin{subfigure}[t]{0.31\textwidth}
        \centering
        \includegraphics[width=\textwidth]{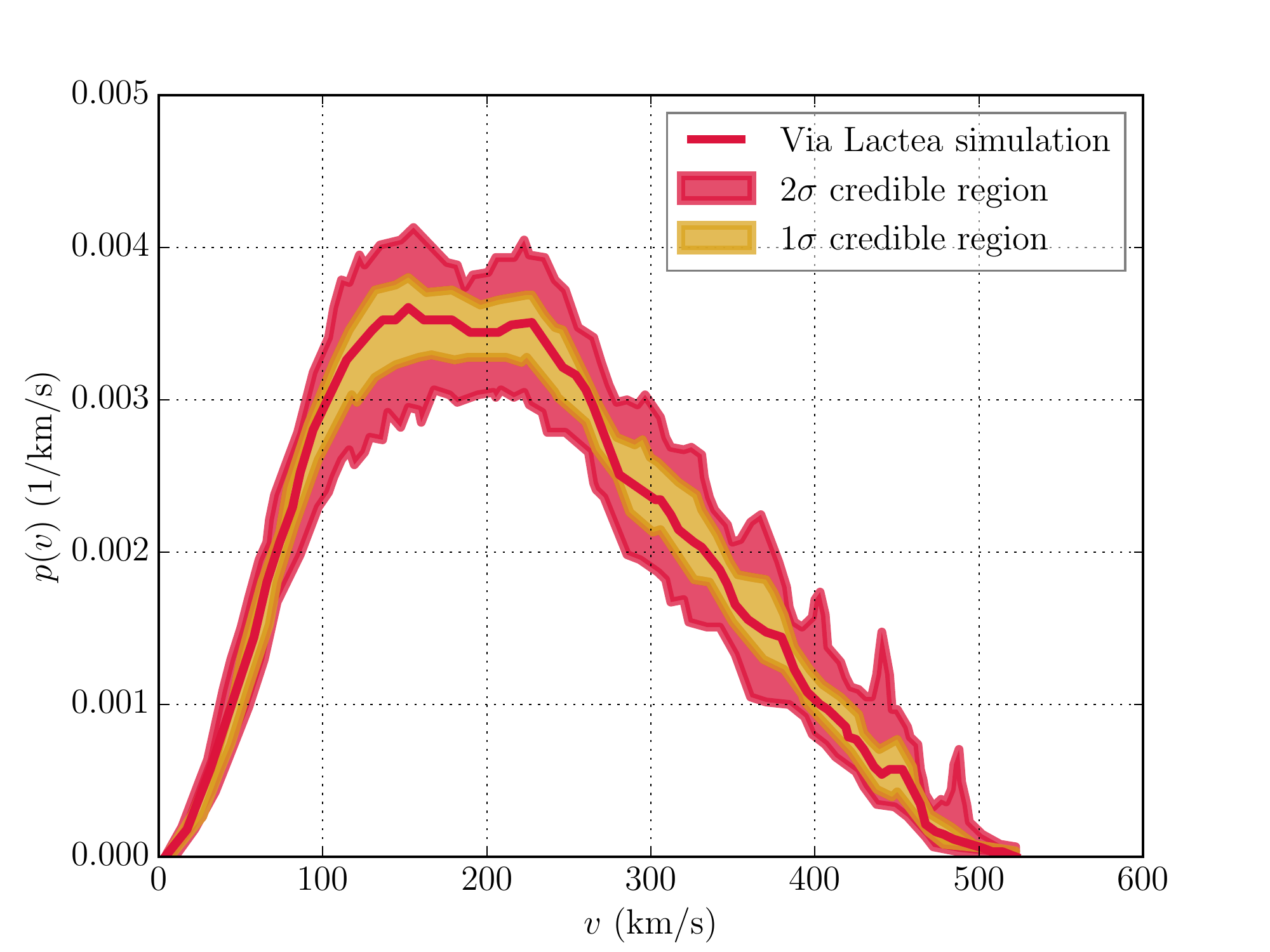}
        \caption{Via Lactea simulations\cite{Kuhlen:2009vh}.}
        \label{fig:simulated}
    \end{subfigure}
    \caption{Velocity distribution and uncertainty from (\subref{fig:mistrust}) truncated Maxwell-Boltzmann distribution with entropic prior reflecting mistrust of background information; (\subref{fig:calibrated}) truncated Maxwell-Boltzmann distribution with entropic prior calibrated calibrated by picking $\beta$ to roughly match uncertainty in simulation; and (\subref{fig:simulated}) simulations of Via Lactea\cite{Kuhlen:2009vh}.}
    \label{fig:calibrate}
\end{figure}

We fix $v_0 =225 \,\text{km/s}$ and $v_\text{esc} = 550\,\text{km/s}$ in $m(v)$ in \refeq{eq:mb} \see{Workgroup:2017lvb}. We could, in principle, treat the escape and modal velocities in the truncated Maxwell-Boltzmann as nuisance parameters and marginalise them or use a numerical form of the profile from simulations in \reffig{fig:simulated} for $m(v)$.  In the former case, however, we note that incidentally, the Maxwell-Boltzmann distribution is itself a maximum entropy distribution. With $m(v_x, v_y, v_z) = \text{const.}$ and a constraint on the expected kinetic energy, $\langle v^2 \rangle = (3\pi - 8)/ (2\pi) v_0^2$, we recover the Maxwell-Boltzmann distribution.\footnote{The peculiar factor in  $\langle v^2 \rangle = (3\pi - 8)/ (2\pi) v_0^2$ simply results from the relation between the modal velocity $v_0$ and the RMS velocity.} We could e.g., fix $m(v_x, v_y, v_z) = \text{const.}$ and marginalise any uncertainty in $\langle v^2 \rangle$.
% The truncated Maxwell-Boltzmann distribution is isotropic. 
% Since the moment functions $w_i(v)$ are angle-averaged, i.e., isotropic, the maximum entropy solutions are isotropic: isotropy is not a further assumption. The only source of anisotropy could be $m(v)$.
We assumed that the profile was isotropic and used angle-averaged moment functions in the galactic rest frame, $w_i(v)$, from \refcite{Gondolo:2017jro}. The moment functions in the galactic rest frame are not isotropic. The distribution $m(v)$ could be a further source of anisotropy, but we picked a truncated Maxwell-Boltzmann distribution, which is isotropic. 

To consider uncertainties, we must switch from a continuum to a discrete number of velocity bins of width $\Delta v$. From a Gaussian approximation of the posterior described in \refcite{Kopec1993}, we find the uncertainty in the profile,
\begin{equation}\label{eq:error_pdf}
(\Delta f(v_i))^2 \approx \sum_j \frac{\hat e_{ij}^2(v_i)}{\lambda_j(v_i)},
\end{equation}
where $\hat e$ and $\lambda$ are the unit eigenvectors and eigenvalues of the positive-definite matrix
\begin{equation}
R_{ij} = \beta \nabla_i\nabla_j S[f, m] + \frac12 \nabla_i\nabla_j \chi^2[f] = \frac{1}{f(v_i)} \beta\delta_{ij} \Delta v + \sum_k \frac{w_k(v_i) w_k(v_j)}{\sigma_k^2} (\Delta v)^2,
\end{equation}
where $\nabla_i = \partial / \partial f(v_i)$. We must stress that this uncertainty vanishes in the limit in which the strength of our prior information diverges, $\beta \to \infty$, since in that case the profile is fixed by the prior information, and that the uncertainty diverges as $\Delta v \to 0$. The latter limit reflects the fact that on microscopic scales $\Delta v \sim 0$ the profile may vary wildly, despite our prior information and since the moments constrain only the macroscopic behaviour of the profile. The calculations of further moments, such as unmodulated signal rates at DAMA/LIBRA, however, reflect macroscopic rather than microscopic structure\cite{gull1999quantified}. Thus the uncertainty in an expectation such as $\langle g \rangle = \sum g(v_i) f(v_i) \Delta v$, found by propagating errors in a linear function,
\begin{equation}\label{eq:error}
(\Delta \langle g \rangle)^2 = g(v_i) R_{ij}^{-1} g(v_j) (\Delta v)^2,
\end{equation}
is finite in the continuum limit $\Delta v \to 0$. 

In \reffig{fig:calibrate} we calibrate $\beta$ by finding $\beta$ such that the uncertainty in the truncated Maxwell-Boltzmann distribution with no moment constraints is similar to the uncertainty in the profile from $N$-body simulations in \refcite{Kuhlen:2009vh}. We estimate the uncertainty at a resolution of about $\Delta v \approx 5\,\text{km/s}$ from the Laplace approximation in \refeq{eq:error_pdf} and find that $\beta \approx 10^4$.

We note that the uncertainty in the simulation results from considering the velocities of about $n = 10^4$ DM particles repeated about 100 times. If the velocities were independently sampled from the same distribution $m(v)$ in each repeat, as in the ``monkey'' argument, we would obtain
\begin{equation}
p(f) \propto e^{n S[f, m]}.
\end{equation}
That is we would \emph{derive} $\beta \approx 10^4$ and its physical interpretation would be the number of DM particles that were sampled to determine $f(v)$. This is not, however, the uncertainty that we wish to describe: we wish to describe uncertainty in the distribution $m(v)$ rather than the statistical uncertainty in an estimate of $m(v)$ from a finite number of samples drawn from $m(v)$. The distribution $m(v)$ is in fact known from simulations with millions of particles and $\beta \gg 10^4$ may be a fairer reflection of uncertainty in $m(v)$ determined from DM only simulations. 

Although our focus is estimating the profile and unmodulated moments for DAMA/LIBRA, we note that we can calculate evidences for the DAMA/LIBRA excess from a Gaussian approximation\cite{BLTJ:BLTJ1338},
\begin{equation}
p(\text{DAMA/LIBRA moments} \given \text{DM, $\beta$}) = \frac{1}{\sqrt{(2\pi)^n}}\frac{1}{\sqrt{\text{det}\,\sigma^2}} \frac{1}{\sqrt{\text{det}\,Z}} e^{\beta S[\hat f, m] -\frac12 \chi^2[\hat f]},
\end{equation}
where $\hat f$ is the maximum entropy solution for that $\beta$ and
\begin{equation}
Z_{ij} = \delta_{ij} + \frac{1}{2\beta} \left.\left(\nabla_i\nabla_k S[f, m]\right)^{-1/2} \left(\nabla_k\nabla_l \chi^2[f]\right) \left(\nabla_l\nabla_j S[f, m]\right)^{-1/2}\right|_{\hat f}.
\end{equation}
This matrix depends on the strength of our prior information, governed by $\beta$, and diverges in the limit $\beta \to 0$, such that the evidence vanishes. This reflects the fact that without prior information the space of possible functions is improper. The ratio of evidences for two models is known as a Bayes factor and reflects the change in relative plausibility of the models in light of data \see{Jeffreys:1939xee}. That is,
\begin{equation}
\frac{
P(\text{DM} \given \text{DAMA/LIBRA moments}, \beta)
}{
P(\text{no DM} \given \text{DAMA/LIBRA moments})
}
=
\frac{
p(\text{DAMA/LIBRA moments} \given \text{DM, $\beta$})
}{
p(\text{DAMA/LIBRA moments} \given \text{no DM})
} \cdot
\frac{
P(\text{DM} \given \beta)
}{
P(\text{no DM})
}.
\end{equation}
where the factor on the right-hand-side is a Bayes factor. We do not specify the ratio of priors for the models and thus do not calculate the posterior odds on the left-hand-side.

\section{Method and results}\label{sec:results}

\begin{table}
\centering
\begin{tabular}{cl}
\toprule
$E$ (keVee) & $S_m$ (1/day/kg/keV)\\
\midrule
$2.25$ & $0.016 \pm 0.0039$\\
$2.75$ & $0.026 \pm 0.0044$\\
$3.25$ & $0.022 \pm 0.0044$\\
$3.75$ & $0.0084 \pm 0.0040$\\
$4.25$ & $0.0110 \pm 0.0036$\\
$4.75$ & $0.0054 \pm 0.0032$\\
$5.25$ & $0.0089 \pm 0.0032$\\
$5.75$ & $0.0039 \pm 0.0031$\\
$6.25$ & $0.00018 \pm 0.0031$\\
$6.75$ & $0.00018 \pm 0.0028$\\
$7.25$ & $0.0015 \pm 0.0028$\\
$7.75$ & $-0.0013 \pm 0.0029$\\
\bottomrule
\end{tabular}
\caption{Noisy measurements of modulated moments from DAMA/LIBRA\cite{Bernabei:2010mq} with $1.17$ tonne-years of exposure reproduced from Tab.~1 of \refcite{Gondolo:2017jro}. The energy bin width was 0.5 keVee.}
\label{tab:sm}
\end{table}

\begin{figure}[t]
    \centering
    \begin{subfigure}[b]{0.475\textwidth}
        \centering
        \includegraphics[width=\textwidth]{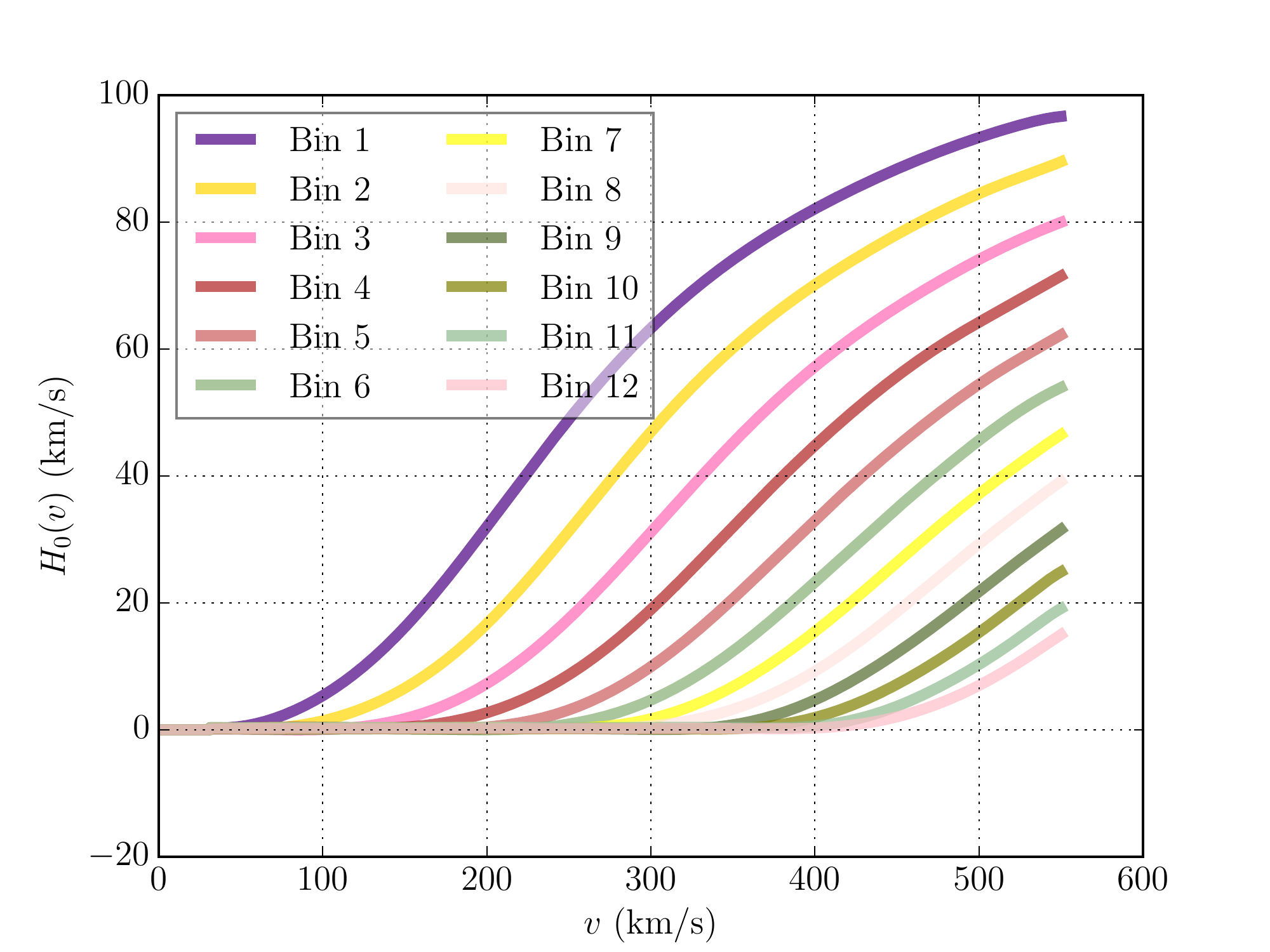}
        \caption{Unmodulated moment functions.}
        \label{fig:H0}
    \end{subfigure}
    \begin{subfigure}[b]{0.475\textwidth}
        \centering
        \includegraphics[width=\textwidth]{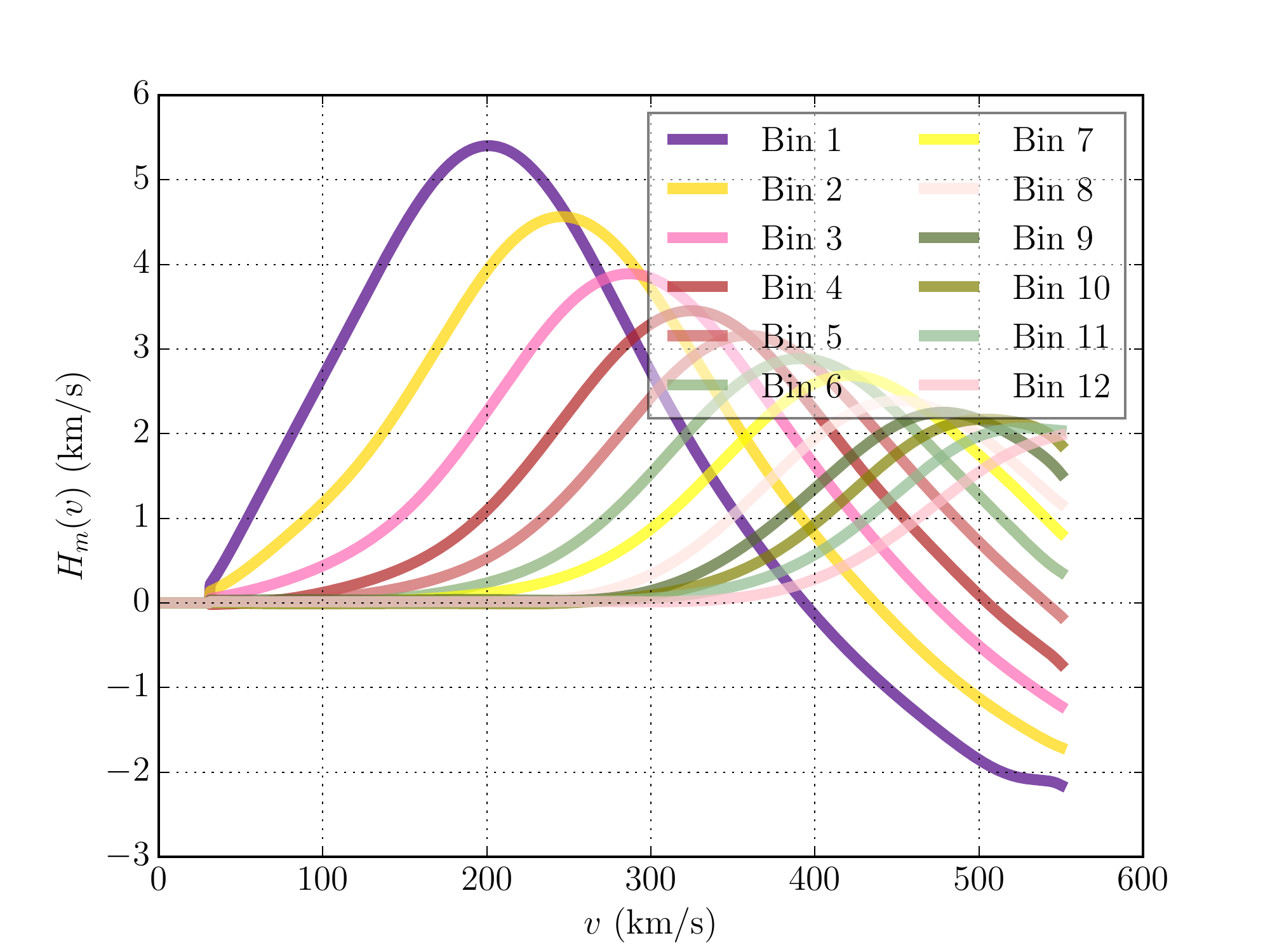}
        \caption{Modulated moment functions.}
        \label{fig:HM}
    \end{subfigure}
    
    \caption{Angle-average galactic rest frame (\subref{fig:H0}) unmodulated and (\subref{fig:HM}) modulated moment functions for the energy bins in \reftable{tab:sm} for a DM mass of $10\gev$ from \refcite{Gondolo:2017jro}. The moment functions must be multiplied by an a priori unknown factor reflecting DM and detector properties, which is profiled in our analysis.}
    \label{fig:moments}
\end{figure}

We focus upon a DM mass of $10\gev$ as favoured by the DAMA/LIBRA anomaly. Our moment functions are angle-averaged modulated ($H_m$) and unmodulated ($H_0$)  detector response functions in the galactic rest frame for spin-independent elastic scattering with sodium in DAMA/LIBRA. The functions were calculated in Fig.~1 and Fig.~2 in \refcite{Gondolo:2017jro} and incorporate all velocity dependence except for that in the profile itself. The moment functions depend upon DM properties, such as its mass, local density and scattering cross section through a single multiplicative factor. We treat this factor as a nuisance parameter. The kernels of the moment functions are plotted in \reffig{fig:moments}. Thus the constraints from the unmodulated moments are
\begin{equation}
S_m^i = k \int_{\mathbf v} H_m^i(v) f(v) dv = \mu_i \pm \sigma_i
\end{equation}
where $S_m^i$ denotes a count rate per day per kg per keV in an energy bin $i = 1, 2, \ldots, 12$; $\mu$ and $\sigma$ are the experimental measurements and uncertainties in \reftable{tab:sm}; and $k$ is the aforementioned multiplicative factor.

\begin{figure}
    \centering
    \begin{subfigure}[t]{0.49\textwidth}
        \centering
        \includegraphics[width=\textwidth]{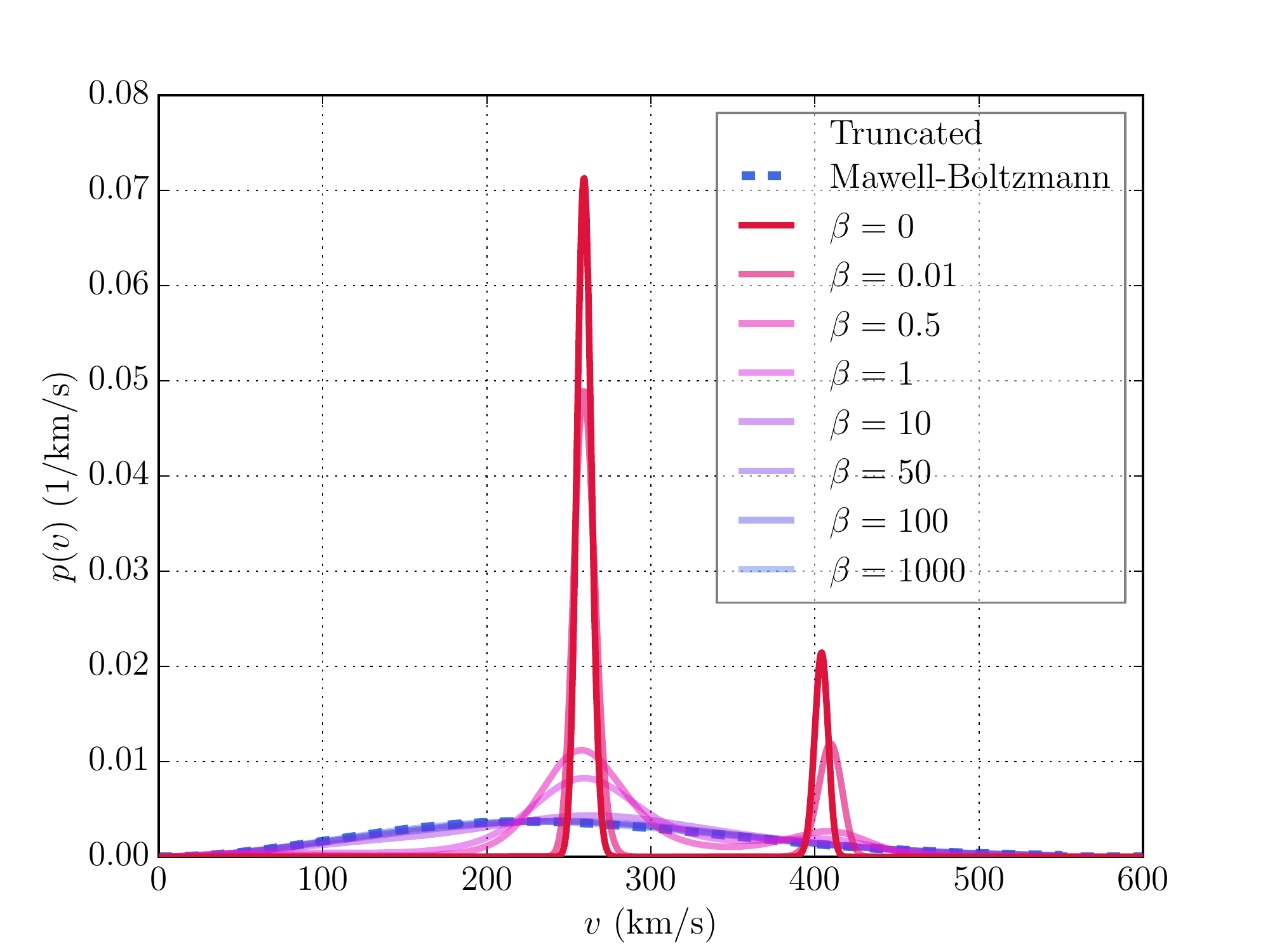}
        \caption{Maximum entropy profiles.}
        \label{fig:max_ent_2d}
    \end{subfigure}
    \begin{subfigure}[t]{0.49\textwidth}
        \centering
        \includegraphics[width=\textwidth]{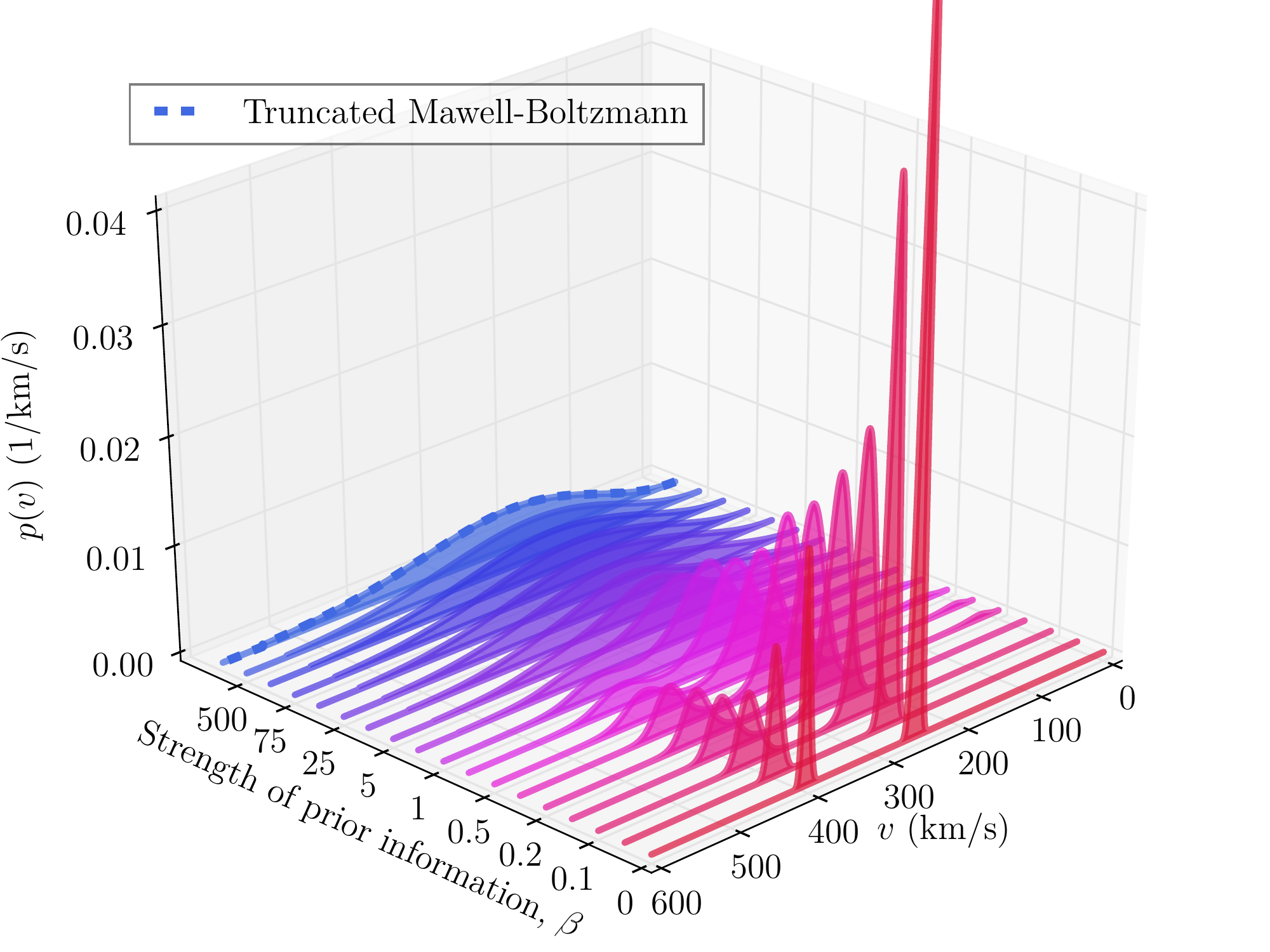}
        \caption{Three-dimensional visualisation of maximum entropy trajectory of profiles.}
        \label{fig:max_ent_3d}
    \end{subfigure}
    
    \caption{Maximum entropy trajectory of profiles for DM with $m_\chi = 10\gev$ found from DAMA/LIBRA measurements of modulated moments (\reftable{tab:sm}). We show the truncated Maxwell-Boltzmann distribution for comparison (dashed blue line). The parameter $\beta$ represents the strength of our conviction that the profile is Maxwellian.}
    \label{fig:max_ent}
\end{figure}

\begin{figure}
    \centering
    \begin{subfigure}[t]{0.48\textwidth}
        \centering
        \includegraphics[width=\textwidth]{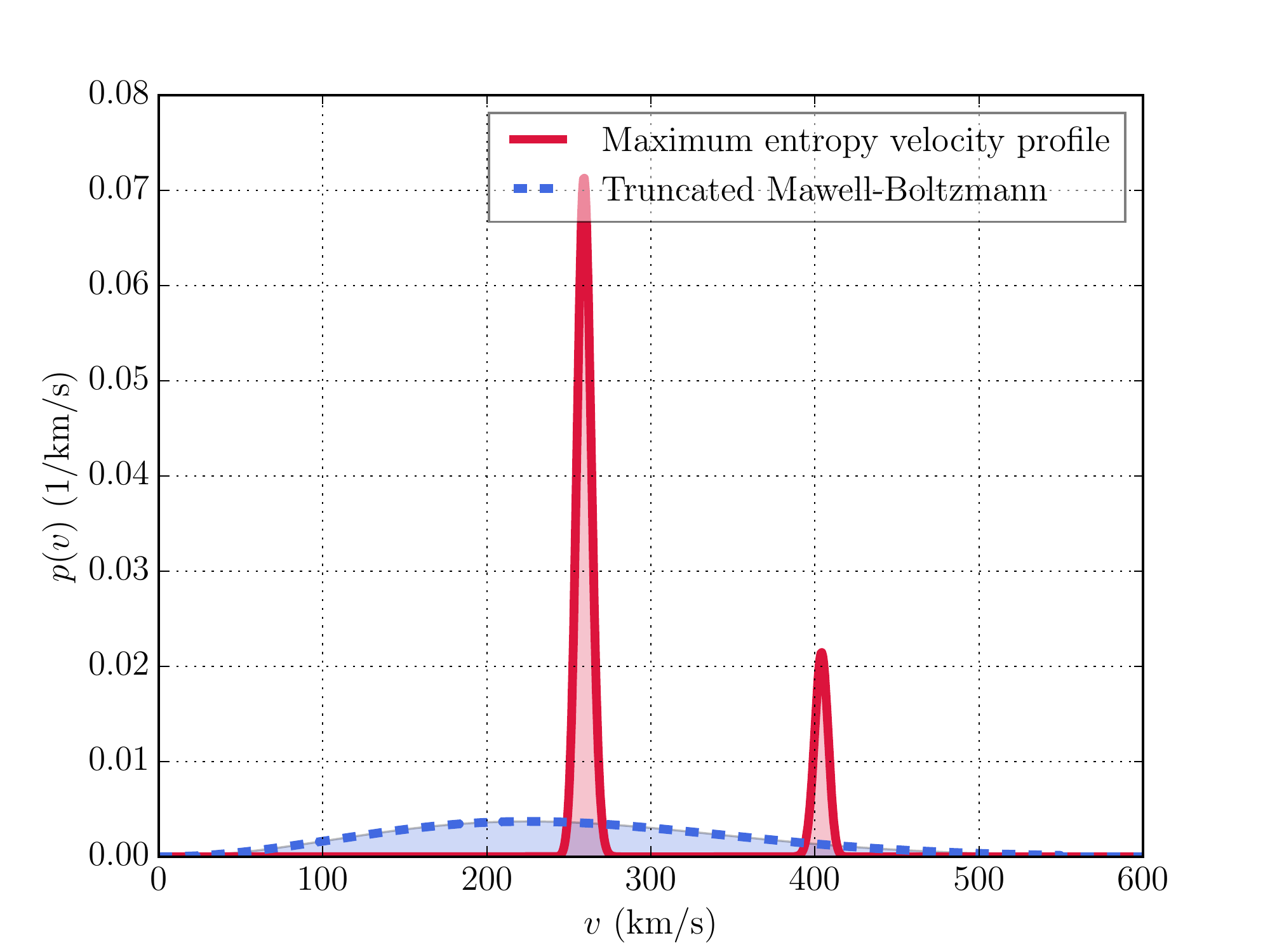}
        \caption{$\beta=0$.}
        \label{fig:max_ent_0}
    \end{subfigure}
    \begin{subfigure}[t]{0.48\textwidth}
        \centering
        \includegraphics[width=\textwidth]{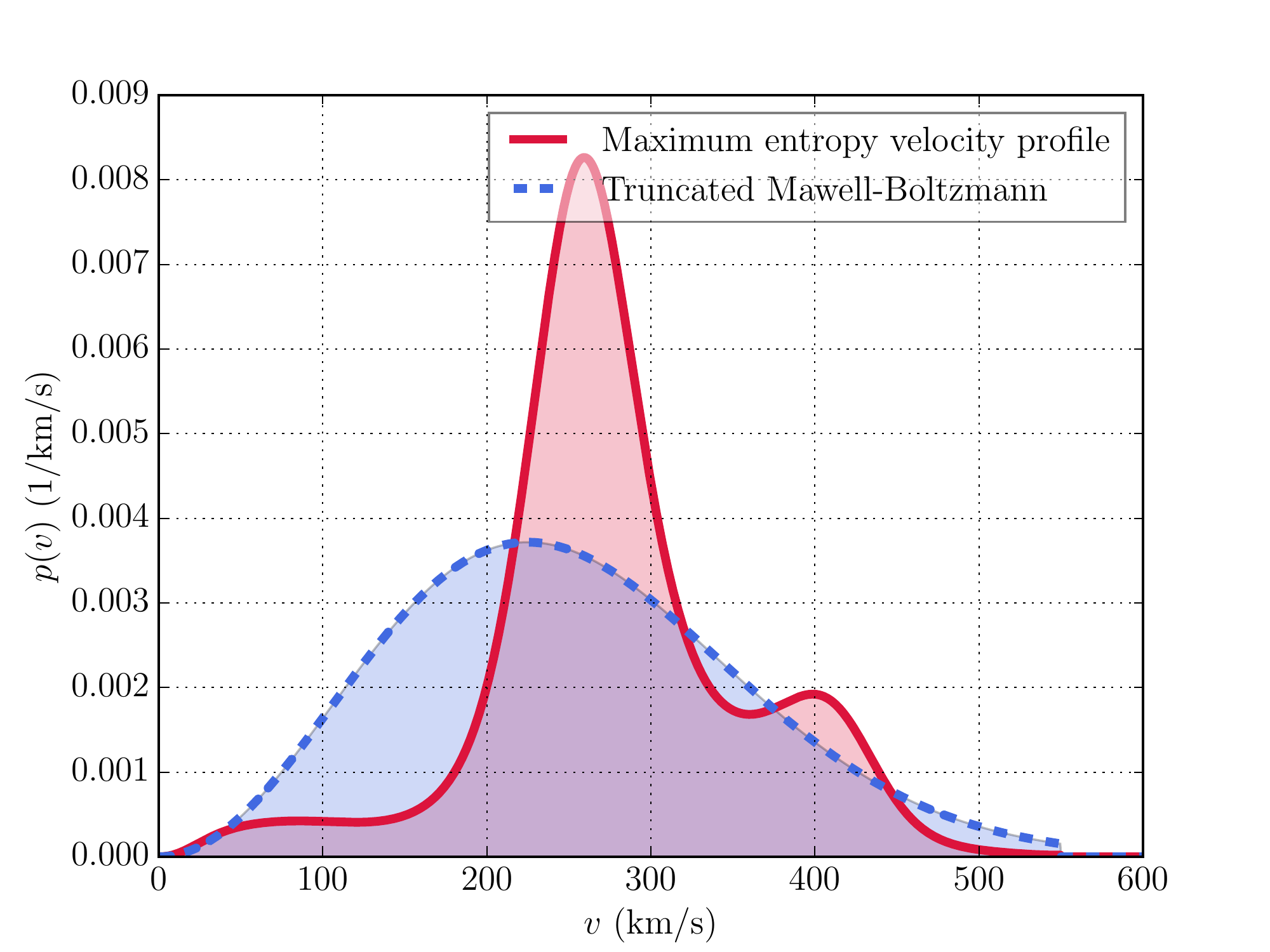}
        \caption{$\beta=1$.}
        \label{fig:max_ent_1}
    \end{subfigure}
    \begin{subfigure}[t]{0.48\textwidth}
        \centering
        \includegraphics[width=\textwidth]{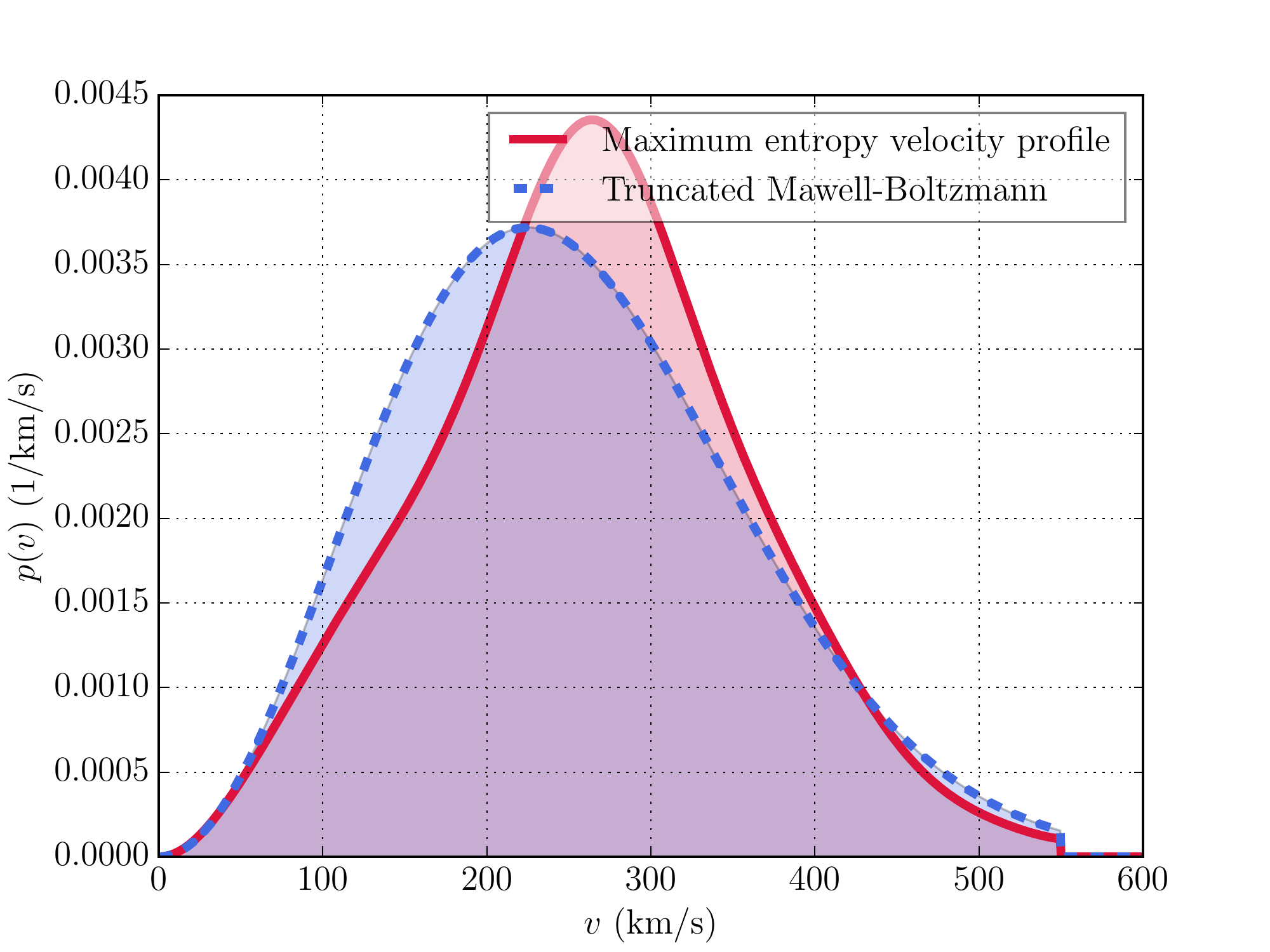}
        \caption{$\beta=10$.}
        \label{fig:max_ent_10}
    \end{subfigure}
    \begin{subfigure}[t]{0.48\textwidth}
        \centering
        \includegraphics[width=\textwidth]{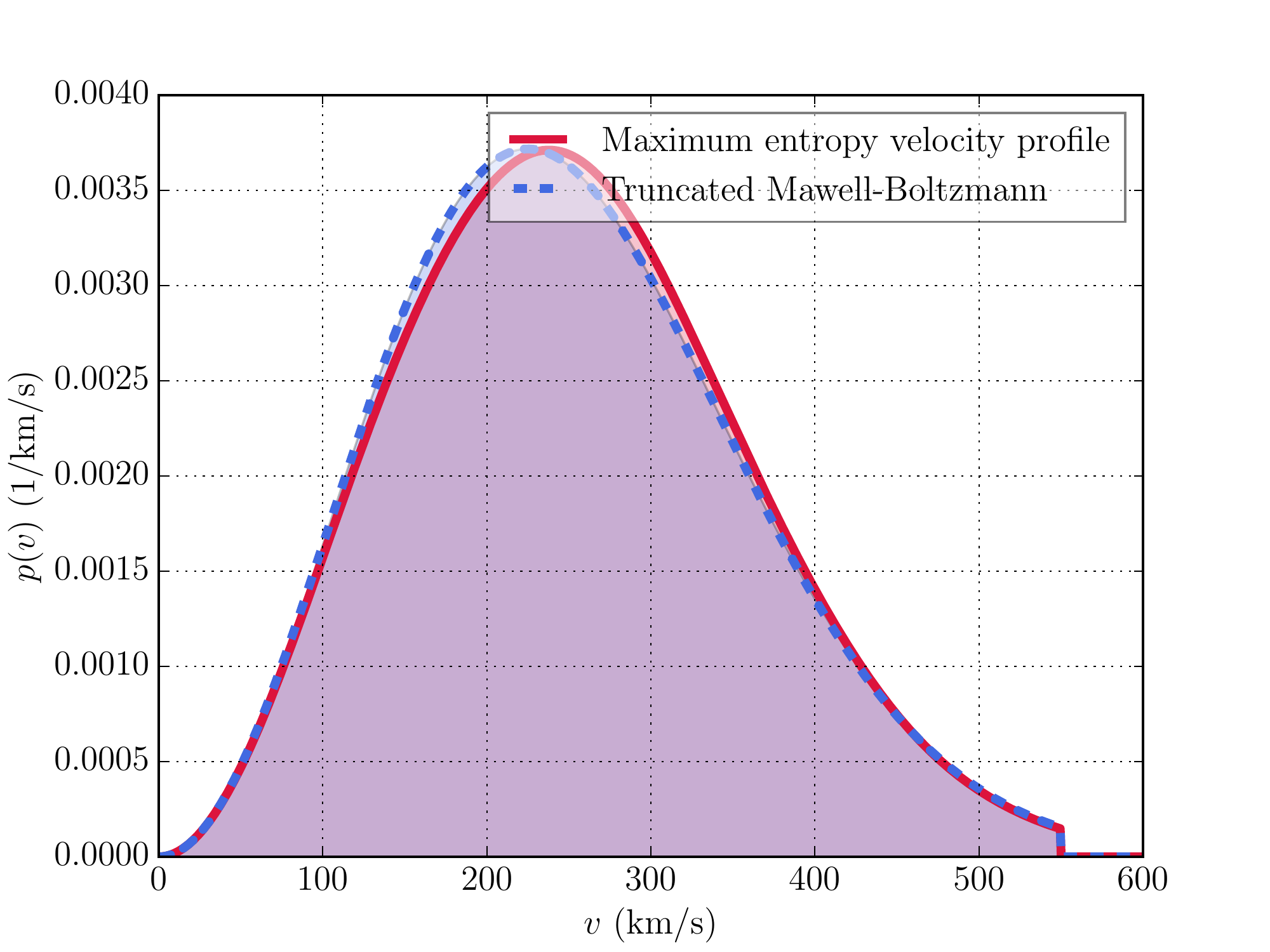}
        \caption{$\beta=100$.}
        \label{fig:max_ent_100}
    \end{subfigure}    
    \caption{Maximum entropy profiles for DM with $m_\chi = 10\gev$ found from DAMA/LIBRA measurements of modulated moments (\reftable{tab:sm}) for (\subref{fig:max_ent_0}) $\beta=0$, (\subref{fig:max_ent_1}) $\beta=1$, and (\subref{fig:max_ent_10}) $\beta=10$ and (\subref{fig:max_ent_100}) $\beta=100$. We show the truncated Maxwell-Boltzmann distribution for comparison (dashed blue line). The parameter $\beta$ represents the strength of our conviction that the profile is Maxwellian.}
    \label{fig:profiles}
\end{figure}

Finding the quantified maximum entropy profiles requires us to maximise \refeq{eq:pdf} with the ansatz in \refeq{eq:sol}. Thus we must maximise a function of twelve Lagrange multipliers that correspond to twelve moment constraints. We fix the multiplicative factor to
\begin{equation}
k = \frac{\sum_i \mu_i S_m^i / \sigma_i^2}{\sum_i (S_m^i / \sigma_i)^2}
\end{equation}
to minimise the chi-squared given the predicted moments. We find the maximum with \texttt{MultiNest}\cite{Feroz:2008xx} and polish it with BFGS and Powell's method\cite{doi:10.1093/comjnl/7.2.155}. We find a minimum $\chi^2=7.03$, which is in reasonable agreement with Fig.~5 in \refcite{Gondolo:2017jro}.  The slight difference may originate from minor differences in the moment functions.

\begin{figure}
    \centering
    \begin{subfigure}[t]{0.49\textwidth}
        \centering
        \includegraphics[width=\textwidth]{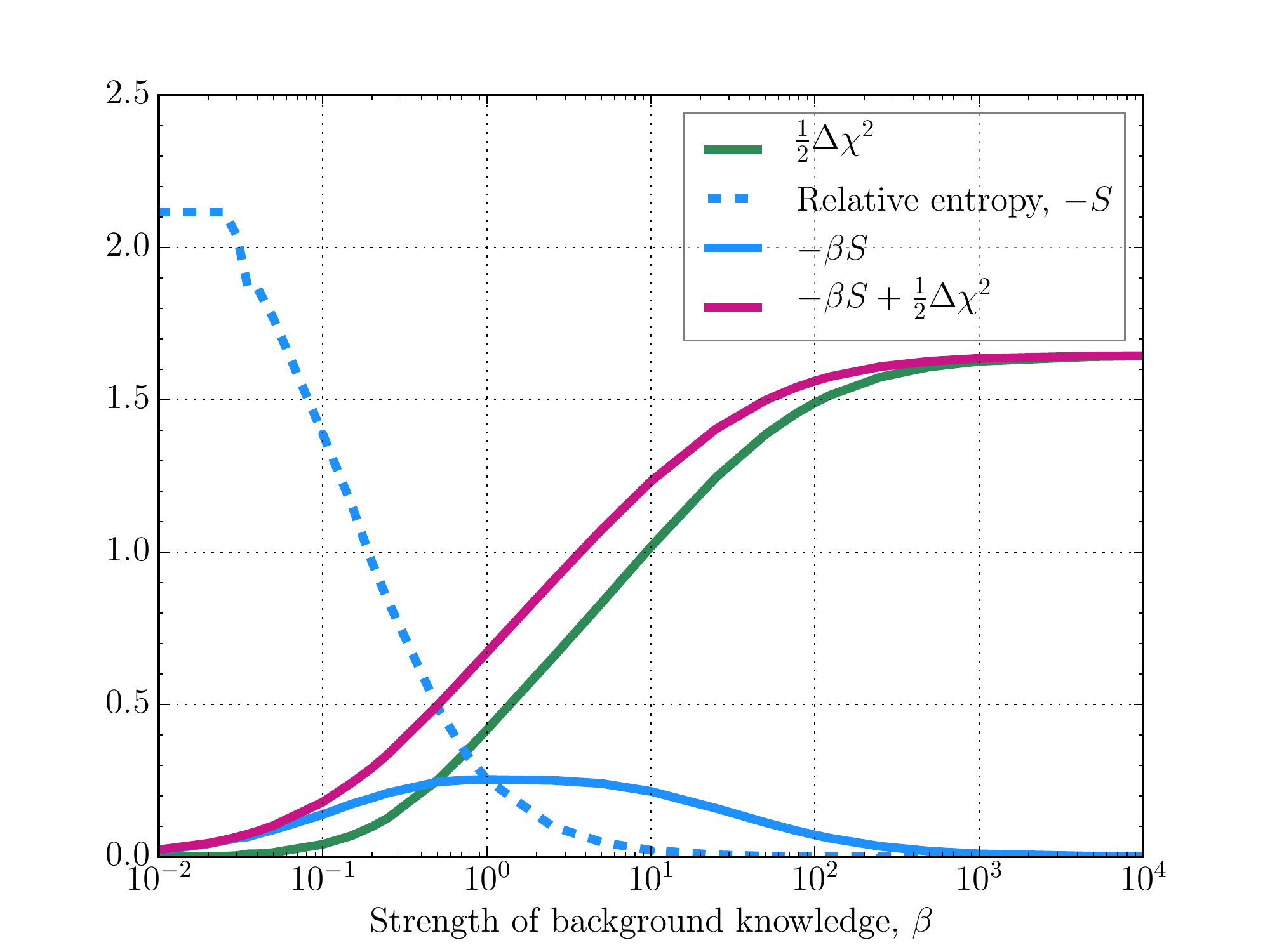}
        \caption{Chi-squared, entropy and their sum as it appears in the posterior as functions of $\beta$.}
        \label{fig:chi_squared}
    \end{subfigure}
    \begin{subfigure}[t]{0.49\textwidth}
        \centering
        \includegraphics[width=\textwidth]{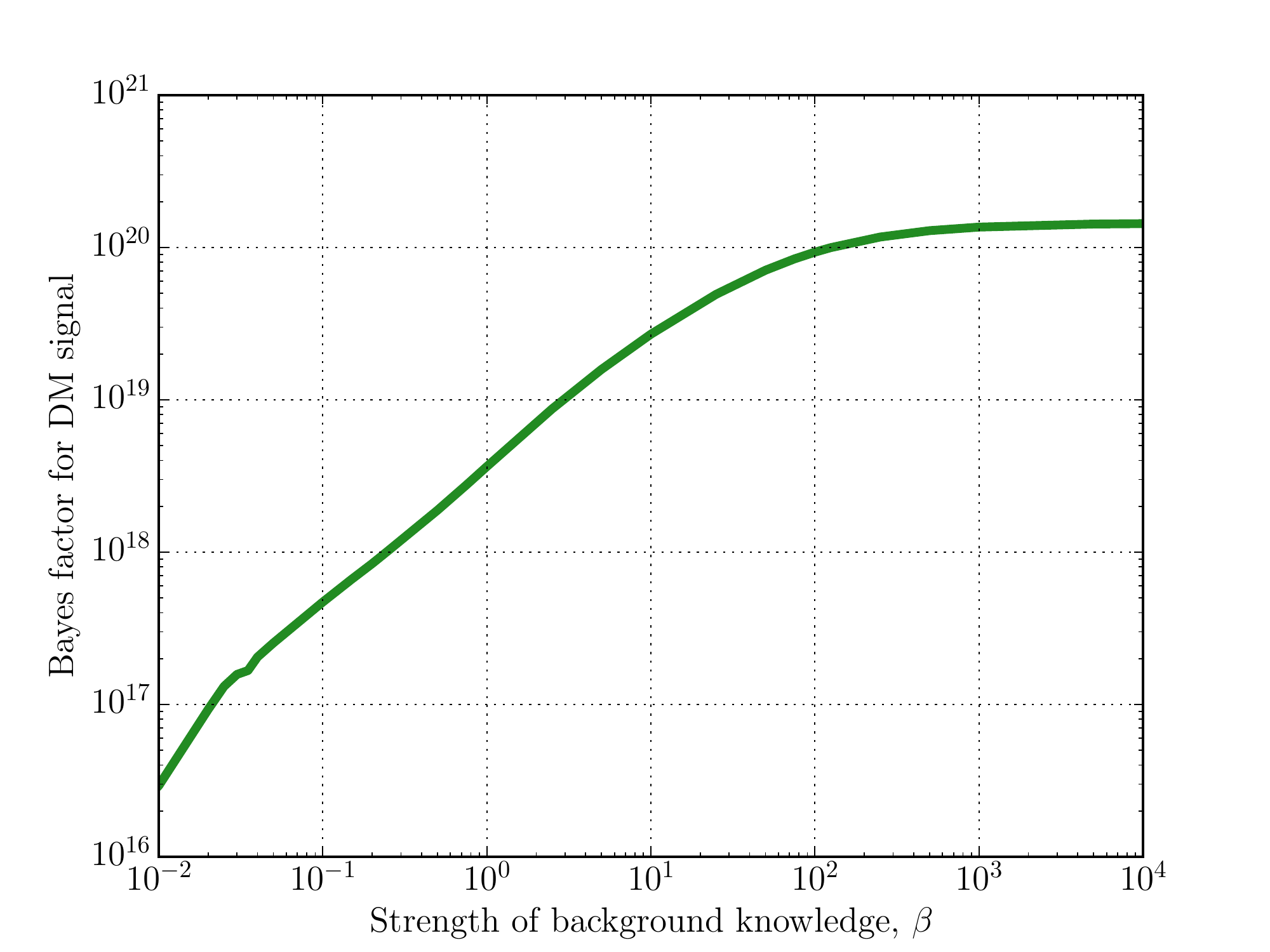}
        \caption{Bayes factor for DM as function of $\beta$.}
        \label{fig:bf}
    \end{subfigure}
    \caption{Maximum entropy trajectories of statistics for model selection as functions of our prior knowledge $\beta$. (\subref{fig:chi_squared}) The interplay between $\Delta \chi^2 \equiv \chi^2 - \min \chi^2$ and the entropy. (\subref{fig:bf}) The Bayes factor for a DM signal versus no DM. The parameter $\beta$ represents the strength of our conviction that the profile is Maxwellian.}
    \label{fig:stat}
\end{figure}

We calculate the maximum entropy profile for many choices of $\beta$ in \reffig{fig:max_ent}. The choice of $\beta$ spans $\beta\approx0$, which discards prior information about the profile, to $\beta \approx 10^4$, which was calibrated to reflect the uncertainty in estimates of the profile from simulations. Intermediate values of $\beta$ reflect a cautious interpretation of simulations and mistrust of information about the profile. The maximum entropy trajectory interpolates between the minimum chi-squared solution at $\beta=0$ and a truncated Maxwell-Boltzmann at $\beta \gg 0$. The minimum chi-squared solution exhibits two sharp modes; a spike at about at about $250\,\text{km/s}$ and a smaller spike at about $400\,\text{km/s}$. 
% This agrees with an analysis in Tab.~1 of \refcite{Ibarra:2017mzt} which shows velocity streams at about $250\,\text{km/s}$ in a Pad\'e approximation. 
As we increase $\beta$, the spikes are smoothed and we rapidly recover a Maxwellian profile. The transition between a spiky profile and a Maxwell-Boltzmann profile occurs fast, and is roughly complete once $\beta \gtrsim 100$. For clarity, we show $\beta=0$, $\beta=1$, $\beta=10$ and $\beta=100$ profiles in \reffig{fig:profiles}. 

We show in \reffig{fig:chi_squared} the interplay between penalties from the entropic prior for deviations from a Maxwell-Boltzmann profile and penalties from the chi-squared for deviations from the DAMA/LIBRA measurements. As the strength of background knowledge increases, the relative entropy between the profile and a Maxwell-Boltzmann vanishes, and the chi-squared increases, as expected. The chi-squared and relative entropy change by only a few units as $\beta \to \infty$, suggesting limited tension between DAMA/LIBRA and background knowledge about the profile. 

In \reffig{fig:bf} we show the entropy trajectory of the Bayes factor from DAMA/LIBRA, that is, the change in relative plausibility of a DM model versus no DM in light of the DAMA/LIBRA anomaly. For $\beta=0$, the DM model is punished by a divergent measure of profiles and thus $B=0$. As we increase $\beta$, the Bayes factor climbs and reaches an asymptote at $B\approx 10^{20}$, suggesting no tension between background knowledge and DAMA/LIBRA. This colossal factor is unsurprising since it is a $9\sigma$ anomaly. The mild dependence upon $\beta$ formalises the fact that the significance of the DAMA/LIBRA anomaly depends on, inter alia, the trust of prior information about the profile. The Bayes factor should, in principle, incorporate factors from marginalising the multiplicative factor in the moment functions and the DM mass. For simplicity and since $B \gg 1$ in any case, we fix them in our calculation.

\begin{figure}[ht]
    \centering
    \includegraphics[width=0.49\textwidth]{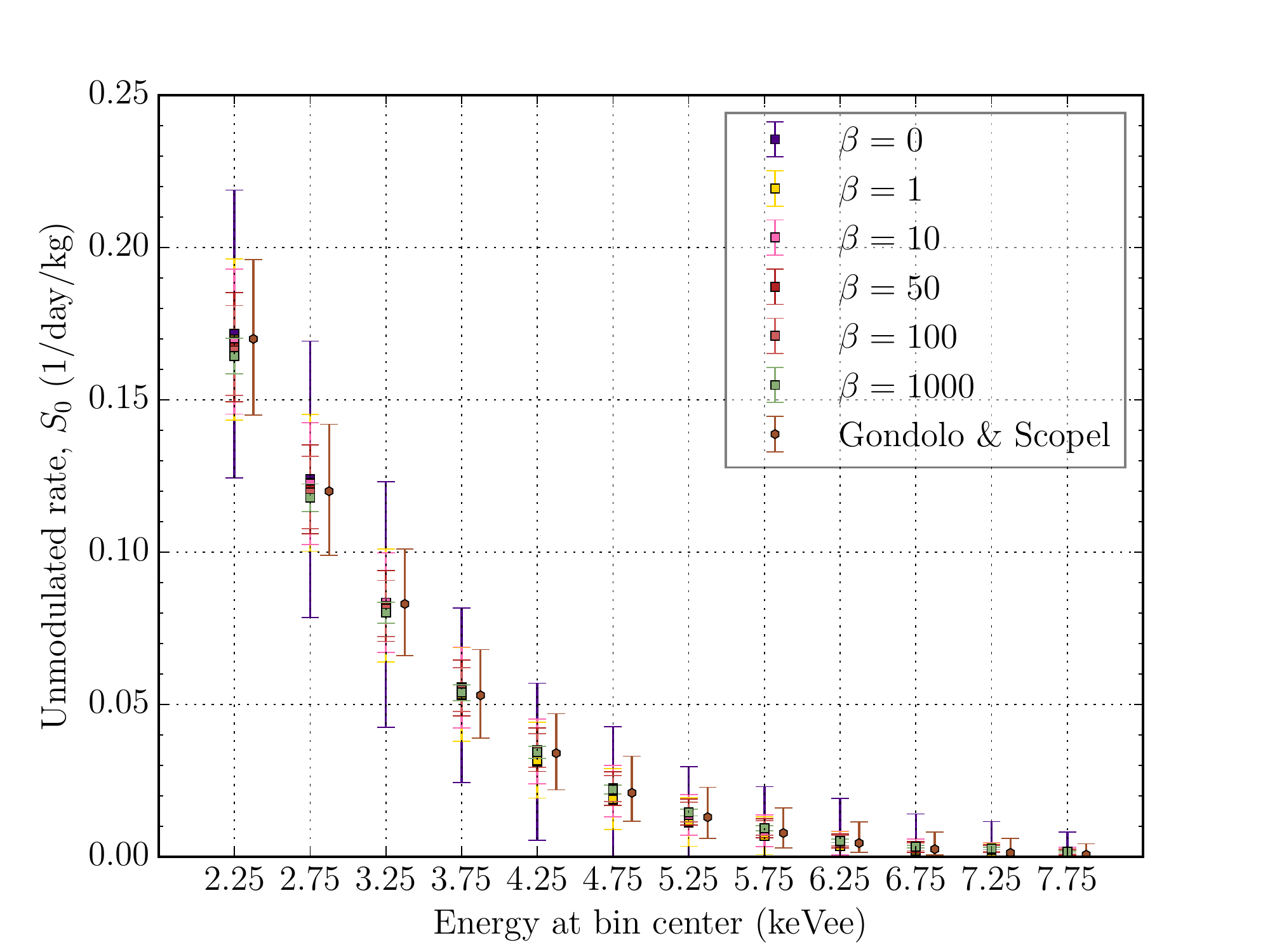}
    \caption{Predictions for unmodulated signal rate, $S_0$, at DAMA/LIBRA for $m_\chi =10\gev$ with an energy bin width of $0.5$ keVee. We show predictions from a frequentist analysis in \refcite{Gondolo:2017jro} for comparison (offset brown hexagons). The parameter $\beta$ represents the strength of our conviction that the profile is Maxwellian.}
    \label{fig:S0}
\end{figure}

Finally, in \reffig{fig:S0} we show the expected modulated signal rates at DAMA/LIBRA and uncertainties from \refeq{eq:error}. They were found analogously to the unmodulated signal rates from
\begin{equation}
S_0^i = k \int_{\mathbf v} H_0^i(v) f(v) dv.
\end{equation}
The predictions are in reasonable agreement with those in \refcite{Gondolo:2017jro}.  Our predictions lie within the frequentist confidence intervals from \refcite{Gondolo:2017jro}, but our uncertainties at $\beta=0$ are greater than the confidence intervals. As we increase $\beta$, the prior information shrinks the confidence interval for the unmodulated moments, as expected. Uncertainties found from  \refeq{eq:error} and  \refeq{eq:error_pdf} may be unreliable for $\beta \approx 0$ as the mode is not a stationary point.

\section{Conclusions}\label{sec:conc}

The non-parametric formalism presented --- quantified maximum entropy --- is novel in DM analysis and allows us to incorporate experimental measurements and background information about the DM profile in a coherent manner, making it particularly appropriate for investigating halo-dependence. We permitted an infinite set of profiles, but profiles far from Maxwellian were punished by an entropic prior. We applied this technique to noisy moment measurements from DAMA/LIBRA, but it could be applied for any data. By varying the parameter $\beta$, which describes the strength of our prior information about the DM profile, we interpolated between a halo-independent and a halo-dependent analysis. The technique has many advantages over Pad\'e approximations, e.g., a coherent treatment and understanding of the role of prior information, weak convergence theorems about maximum entropy and the fact that the estimate for the profile is not a sum of discrete velocity streams. We assumed an isotropic profile in the galactic rest frame and considered angle-averaged moments from DAMA/LIBRA, but could, in principle, consider anisotropic likelihoods and profiles with quantified maximum entropy.

As an example, we focused upon the DAMA/LIBRA anomaly with DM mass of $10\gev$. As we increased the strength of our conviction that the profile was Maxwellian, the preferred profile changed smoothly from a spiky profile with peaks at about $250\,\text{km/s}$ and $400\,\text{km/s}$ to a Maxwell-Boltzmann distribution. For any $\beta \gtrsim 1$, the Bayes factor from DAMA/LIBRA favoured DM by about $10^{20}$. There was limited tension between DAMA/LIBRA data and a Maxwellian profile; the chi-squared and relative entropy changed by a few units as $\beta$ was increased and the Bayes factor, in fact, increased as $\beta$ increased.

We predicted the unmodulated moments $S_0$ at DAMA/LIBRA. The predictions were closely clustered for different $\beta$, suggesting that background information plays a limited role in predictions for unmodulated moments at DAMA/LIBRA. However, the uncertainties in the predicted moments were sensitive to $\beta$: without prior information there was a substantial uncertainty in predictions for $S_0$. This may be an overestimate resulting from a Gaussian approximation. 

Since the method is novel, a few questions remain. In particular, the Gaussian approximation of the posterior may require further investigation, and, if it is inadequate, an improved technique for estimating uncertainties in further moments may be required. It may be necessary to bin the velocity into $N$ bins and directly calculate the posterior probability of $f(v_i)$, thus avoiding Gaussian approximations. Furthermore, the DM mass and properties should be marginalised upon a particular prior, rather than profiled as in this analysis. Once such details are chosen, it remains to perform a global analysis incorporating data from all DM experiments, including conflicting results from e.g., XENON and LUX. This may present further difficulties, e.g., finding an ansatz for the quantified maximum entropy solution as in \refeq{eq:sol}.

\section*{Acknowledgements}

This work in part was supported by the ARC Centre of Excellence for Particle Physics at the Terascale.

\end{document}